\documentstyle[pre,preprint,aps,epsfig]{revtex} 
\tighten

\begin{document} 
\draft

\title{ The Potts frustrated model: relations with glasses
}  
\author{
Giancarlo Franzese$^{1,3}$~\thanks{Present address: Center for Polymer
Studies, Department of Physics, Boston University, 590 Commonwealth
Avenue, Boston, MA 02215} 
and Antonio Coniglio$^{2,3}$
}
\address{
$^1$ Dip. di Fisica ``E.Amaldi'', Universit\`a Roma Tre,
via della Vasca Navale 84, 00146 Roma, Italy\\
$^2$ Dip. di Scienze Fisiche,
Universit\`a di Napoli, Mostra d'Oltremare Pad.19, 80125 Napoli
Italy\\
$^3$ Istituto Nazionale per la Fisica della Materia, UdR Napoli
}

\maketitle 

\begin{abstract}
Similarities between fragile glasses 
and spin glasses (SG) suggest the study of frustrated spin 
model to understand the complex dynamics of glasses above the glass transition.
We consider a frustrated spin model with Ising spins and 
$s$-state Potts spins both with and without 
disorder. We study the two models by Monte Carlo simulations in two
dimensions.
The Potts spins mimic orientational degrees of freedom and the coupled
frustrated Ising spins take into account for frustrating effects like
the geometrical hindrance. 
We show that in this model dynamical transitions and
crossovers are related to static transitions.
In particular, when disorder is present, as predicted and verified in
SG, a dynamical transition between
high-temperatures exponential to 
low-temperatures  non-exponential correlation functions
numerically coincides with the ordering temperature
of the  ferromagnetic regions, i.e. the Griffiths temperature $T_c(s)$,
while a crossover 
between power law growth of correlation times to Arrhenius law
occurs near a Potts transition at $T_p(s)<T_c(s)$.
In the model without disorder, where $T_c(s)$ is not defined, 
both dynamical transition and crossover occur at $T_p(s)$.
Furthermore the static susceptibility and the
autocorrelation times of quantities depending on Potts spins diverge 
at $T_p(s)$. 
This is reminescent of recent experimental results on glass-forming liquids.

\end{abstract}

\section{Introduction}

Glassy systems, like  supercooled liquids,
polymers, granular material, vortex glasses, 
ionic conductors, colloids, plastic crystal  and spin glasses, 
can be defined as materials characterized by a
slowing down of one or more degrees of freedom 
which prevents the system to reach the equilibrium as the temperature
decreases (Angell 1984, G\"otze 1989, Angell 1995, Ediger {\it et al.} 1996).

In particular, spin glasses (SG) are 
dilute alloys of magnetic atoms in low concentration in a non-magnetic 
matrix where the relevant degrees of freedom are the impurities spins 
interacting via random magnetic couplings. 
The randomness gives rise to {\em frustration}, 
i.e. a competition that prevents
the systems to minimize the global free energy (Mezard {\it et al.} 1987). 

One of the characteristic phenomenon of glassy systems is their
complex dynamics well above the calorimetric glass transition at
temperature $T_g$. 
For example there is a dynamical transition
at $T^*>T_g$  from high-temperatures 
exponential to low-temperatures non-exponential correlation
functions. Another feature is the crossover at $T_1$ with $T_g<T_1<T^*$
between a high-temperature
power  law growth of correlation times 
to an Arrhenius behaviour $\tau\sim \exp(\Delta E/k_BT)$, where $\Delta
E$ is the activation energy. This phenomenology is well
seen in experiments 
(Angell 1984, Angell 1995, Ediger {\it et al.} 1996
Andreozzi {\it et al.} 1998). 

Here we show the numerical results on frustrated models with and
without disorder, the Potts SG (PSG) and the Potts fully frustrated (PFF)
model respectively, where the dynamical anomalies correspond to static
transitions, giving some insights on the relation between dynamics and
statics in glassy systems and 
clarifying the role played by frustration and by disorder.
The model could be also useful to understand some 
recent experimental results on finite-temperature diverging static
susceptibility in glassy systems (see for example 
Dixon {\it et al.} 1990, Menon and Nagel 1995, Leheny and Nagel 1997).

\section{The PSG  and PFF models}

Let us consider the Hamiltonian
\begin{equation}
H=-sJ\sum_{\langle i,j \rangle} [\delta_{\sigma_i \sigma_j}
(\epsilon_{i,j}S_i S_j+1)-2]
\label{hamiltonian}
\end{equation}
where to each lattice site is associated an Ising spin $S_i=\pm 1$ and a 
$s$-state Potts spin $\sigma_i=1, \dots , s$.
The sum is extended over all nearest neighbor (NN) sites,
$\epsilon_{i,j}=\pm 1$ is the sign of interaction and
$J$ is the strength of interaction.
When the $\epsilon_{i,j}$ are quenched random variables we will refer
to the model as the PSG, while when they are distributed in a 
deterministic way to frustrate each lattice cell we will refer to it as
the PFF.
For $s=1$ the PSG (PFF) model recovers the Ising SG (FF) model.

The model can be interpreted as a system with $s$-states orientational
degrees of freedom frustrated by means of coupled Ising spins which take
into account for effects due to geometrical hindrance (e.g. like in 
plastic crystals, where
the centers of mass of the molecules form a regular crystal but the
molecules are frustrated respect to the orientational degrees of
freedom, or in o-terphenyl or in glycerol).
The way in which the frustration is introduced distinguishes this model
by other models, e.g. the Potts glass (Kirkpatrick and Wolynes 1987,
Thirumalai and Kirkpatrick 1998),  used to study
structural glasses.

Both PSG and PFF models can be mapped on a Fortuin-Kasteleyn (FK)
percolation model 
(Fortuin and Kasteleyn 1972, Coniglio and Klein 1980), 
defining the FK clusters as maximal sets of spins connected by
bonds activated between NN spins with probability
$p=1-e^{-2sJ/k_BT}$
($k_B$ is the Boltzmann
constant)
when both the NN Potts and Ising spins minimize the coupling energy and using 
the relation
\begin{equation}
Z\{\epsilon_{i,j}\}=\sum_{\{S_i,\sigma_i\}}e^{-H/k_BT}=\sum_C W_s(C)
\label{partition_function}
\end{equation}
where $W_s(C)=0$ if the cluster configuration $C$ 
contains any frustrated loop (defined below), otherwise
$W_s(C)=p^{|C|}(1-p)^{|A|}(2s)^{N(C)}$
where $N(C)$ is the number of clusters in 
the configuration $C$, $|C|$ is the number of activated bonds and
$|C|+|A|$ is total  
number of interactions. A frustrated loop is a closed path that
has an odd number of antiferromagnetic interactions 
(Coniglio {\it et al.} 1991).
While the Hamiltonian (\ref{hamiltonian}) is defined only for integer
$s>1$,  the partition function (\ref{partition_function}) is meaningful for
any values of $s$ (Fierro {\it et al.} 1999).

\section{Statics and dynamics of the models}

The numerical phase diagrams of PSG and PFF model are shown 
in (Franzese and Coniglio  1998, Franzese 1999)
 and are qualitatively reproduced in
figure \ref{fig1} for the 3D case. Two transition temperatures are seen. 
The lower temperature $T_{SG}(s)$ (or $T_{FF}(s)$) corresponds to a SG
(or FF respectively)
transition in the universality class of $\pm J$ Ising SG (or Ising FF) model.
The higher $T_p(s)$ corresponds to the
percolation transition of FK clusters and for any integer $s>1$ 
marks a real thermodynamic transition in the universality class
of a ferromagnetic $s$-state Potts model (Wu 1982). 
At $T_p(s)$ the susceptibility of Potts spins
diverges (Franzese and Coniglio  1998, Franzese 1999). 
In the PSG model, above $T_p(s)$, there is the  Griffiths temperature
$T_c(s)$ defined as the critical temperature of a ferromagnetic model with
the same number of states of the disordered model ($2s$ for the
$s$-state PSG model). This transition is consequence of the presence of
ferromagnetic regions, due to disorder, and vanishes for vanishing external
field (Griffiths 1969).

From the dynamical point of view we have studied 
(Franzese and Coniglio 1999, Franzese 1999) for $s=2$
both for PSG and PFF the time dependent nonlinear susceptibility
\begin{equation}
\chi_{SG}(t)=\frac{1}{N} \overline{\left\langle \left[\sum_i
S_i(t+t_0) S_i(t_0)\right]^2\right\rangle}
\label{chi_sg_t}
\end{equation}
(where $N$ is the total number of spins and $t_0$ is the equilibrium time)
that converges asymptotically for $t\rightarrow \infty$ to the usual static
nonlinear susceptibility.
The normalized correlation function is
\begin{equation}
f_\chi=\frac{\chi_{SG}(t)-\chi_{SG}(t=\infty)}{\chi_{SG}(0)
-\chi_{SG}(t=\infty)}
\end{equation}
with $\chi_{SG}(0)=N$.
Following (Campbell and Bernardi 1994)
 the infinite size behavior of $f(t)$ has 
been extrapolated at every $t$ plotting the data for finite linear sizes 
$L=20$, 25, 30, 40 vs $1/L$. 

To test the form of $f(t)$ we fitted the data $i)$ with a 
simple exponential, finding good fits only asymptotically for long time
and for high temperatures, $ii)$ with a stretched exponential function
$f_0\cdot \exp[(t/\tau)^\beta]$, 
finding that it fails to fit the data only for short times, $ii)$ with
the form 
$f_0\cdot t^{-x} \exp[(t/\tau)^\beta]$
suggested by Ogielski (Ogielski 1985), finding 
that it fits very well the data over all the time's range and the
temperature's range (Franzese and Coniglio 1999).

The parameters $\beta$, $\tau$ used in the fit $ii)$ and $iii)$ for
$\chi_{SG}$ are 
plotted in figures \ref{parpsg}, \ref{parpff} for PSG and PFF
respectively. In the same figures is shown also 
the integral correlation time defined as
\begin{equation}
\tau_{int}=
\lim_{t_{max}\rightarrow \infty}\frac{1}{2}+\sum_{t=0}^{t_{max}}f(t)
\label{int}
\end{equation}
where $f$ is the generic correlation function.

In figure \ref{parpsg} for PSG it is possible to distinguish three dynamical
regions. A high temperatures region for $T>T^* \simeq T_c(s)$ where the
correlations functions are exponential ($\beta=1$); an intermediate region for
$T^*>T >T_1 \simeq T_p(s)$ where $\beta<1$ and the correlation time's growth
is a power law forecasted by the Mode Coupling Theory (MCT)
(G\"otze 1989) or mean field (MF) theory 
(see for example Bouchaud  {\it et al.} 1998 and references therein);
a lower temperatures region for $T< T_1$ where
activated processes dominate the dynamics (i.e. the correlation times
follow an Arrhenius law) as in the experimental glasses
(Angell 1984, Angell 1995, Ediger {\it et al.} 1996,
Andreozzi {\it et al.} 1998). 
By the way, these results
show that the prediction $T^*=T_c$ given for the SG (Randeria {\it et
al.} 1986, Cesi {\it et al.} 1997) is
valid also in this model.

In figure \ref{parpff} for PFF, where $T_c(s)$ is not defined for the
absence of disorder, the intermediate region disappears and both the
dynamical transition at $T^*$ and the crossover at $T_1$ occur at a
temperature numerically consistent with $T_p(s)$.

For the PFF model we have calculated also the 
correlation functions for the Potts order parameter
$M=[s ~\mbox{max}_i(M_i)-1]/(s-1)$
(where $M_i$ is the density of Potts spins in the
$i$-th state) defined as
\begin{equation}
f_M(t)=\left[\frac{\langle \delta(t+t_0) \delta(t_0) \rangle}{\langle
\delta(t_0)^2 \rangle }\right] ,
\label{corrM}
\end{equation}
(where $\delta(t)=M(t)-\langle M\rangle$ and $t_0$ is the equilibration time).
At the Potts transition temperature $T_p(s)$ the static susceptibility
related to $M$ diverges (Franzese 1999) and the parameter $\beta$ and $\tau$ 
for the fit forms $ii)$ and $iii)$ of $f_{M}$ becomes less than one and 
diverging respectively, as shown in figure \ref{pff_M}. The divergence
of $\tau$ comes out from the observation that $\tau(T,L)$ grows for
increasing $L$ at any $T$ and there is a cusp at a temperature
consistent to $T_p(s)$ within the numerical error.

\section{Conclusions}

PSG and PFF model reproduce the experimental phenomenology of
fragile glasses  above 
the glass transition, recovering the dynamical transition between
exponential and non-exponential correlation functions and the crossover
between the power law growth, predicted by MF theories and MCT, and the 
activated Arrhenius regime for the correlation times.
In particular, the disordered PSG model shows three different dynamical
regimes experimentally seen for spin $1/2$ probes in glass-forming liquids
like o-terphenyl (Andreozzi {\it et al.} 1998). 

The models show that these dynamical transitions and crossovers can be
related to static thermodynamic transitions: in the PSG 
to the Griffiths transition and to  the Potts transition. while in the
PFF model only to the Potts transition. 
The mapping of these models on FK percolation model, show
that the Potts transition coincides with a percolation transition where
frustrated loop are excluded, giving a geometrical interpretation of the
effects of frustration.

Furthermore at the Potts transition 
the susceptibility associated to the Potts variables diverges. This
result is reminescent of recent
experimental findings
(see for example 
Dixon {\it et al.} 1990, Menon and Nagel 1995, Leheny and Nagel 1997).
In the models presented here this
divergence is explained by the ordering of the orientational degrees of 
freedom.
If such a divergence is present also in structural glasses like plastic
crystals is matter of debate (Schneider {\it et al.} 1999).

Moreover the PFF model emphasizes that dynamical anomalies are not
related to the presence of frustration {\em and} disorder, but
also to the sole frustration and in this case $T_p$, that is the
ordering temperature of the orientational degrees of freedom, marks the onset
of these anomalies.
This could be relevant in the study of dynamical behavior of
experimental FF systems like Josephson junction arrays.

~\\


\noindent
Andreozzi,~L., Giordano,~M., and Leporini,~D., 1998, {\it
J. Non-Cryst. Solids}, {\bf 235-237}, 219.

\noindent
Angell,~C.A, 1984, {\it Relaxation in Complex Systems}, edited by
Ngai,~K.L, and Wright,~G.B. (Washington: Office of Naval Research).

\noindent
Angell,~C.A, 1995, {\it Science}, {\bf 267}, 1924.

\noindent
Bouchaud,~J.P., Cugliandolo,~L.F., Kurchan,~J., Mezard,~M., 
1998. {\it Spin Glasses and
Random Fields}, edited by Young,~A.P., (Singapore: World Scientific).

\noindent
Campbell,~I.A., and Bernardi,~L., 1994, {\it Phys. Rev. B}, {\bf 50},
12~643.

\noindent
Cesi,~F., Maes,~M., and
Martinelli,~F., 1997, {\it Commun. Math. Phys.}, {\bf 188}, 135.

\noindent
Coniglio,~A., di Liberto,~F., Monroy,~G., and Peruggi,~F. 
1991, {\it Phys. Rev. B}, {\bf 44}, 12~605.

\noindent
Coniglio,~A., and Klein,~W., 1980, {\it J. Phys. A}, {\bf 13}, 2775.

\noindent
Dixon,~P.K., Wu,~L., Nagel,~S.R., Williams, and J.P.~Carini, 
1990, {\it Phys. Rev. Lett.}, {\bf 65}, 1108.

\noindent
Dixon,~P.K., Wu,~L., Nagel,~S.R., Williams, and J.P.~Carini,  1991, 
{\it Phys. Rev. Lett.}, {\bf 66}, 960.

\noindent
Ediger,~M.D, Angell,~C.A, and Nagel,~S.R., 1996, {\it J. Phys. Chem},
{\bf 100}, 13~200.

\noindent
Fierro,~A., Franzese,~G., de Candia,~A., and Coniglio,~A., 1999, {\it
Phys. Rev E}, {\bf 59}, 60.

\noindent
Fortuin,~C.M., and Kasteleyn,~P.W., 1972, {\it  Physica}, {\bf 57}, 536.

\noindent
Franzese,~G., 1999, preprint.

\noindent
Franzese,~G., and Coniglio,~A., 1998, {\it Phys. Rev. E}, {\bf 58}, 2753.

\noindent
Franzese,~G., and Coniglio,~A., 1999, preprint 
http://xxx.lanl.gov/abs/cond-mat/9803043
in press on {\it Phys. Rev. E}, {\bf 59}, vol. 6.

\noindent
G\"otze,~W., 1989, {\it Liquids, freezing and glass transition}, 
edited by Hansen,~J.P., Levesque,~D., Zinn-Justin,~J. (Amsterdam: North
Holland).

\noindent
Griffiths,~R.B., 1969, {\it Phys. Rev. Lett.}, {\bf 23}, 17.

\noindent
Kirkpatrick,~T.R., and Wolynes,~P.G., 1987, {\it Phys. Rev. B}, {\bf 35}, 3072.

\noindent
Leheny,~R., and Nagel,~S.R., 1997, {\it Europhys. Lett.}, {\bf 39}, 447.

\noindent
Menon,~N., and Nagel,~S.R, 1995, {\it Phys. Rev. Lett.}, {\bf 74}, 1230.

\noindent
Mezard,~M., Parisi,~G., and Virasoro,~M.A., 1987, {\it Spin Glass Theory
and Beyond} (Singapore: World Scientific).

\noindent
Ogielski,~A.T., 1985, {\it Phys. Rev. B}, {\bf 32}, 7384.

\noindent
Randeria,~M., Sethna,~J.P., and Palmer,~R.G., 1986, {\it
Phys. Rev. Lett.}, {\bf 57}, 245.

\noindent
Schneider,~U., Brand,~R., P.~Lunkenheimer, and A.~Lloidl, 1999,
preprint http:\-//xxx.\-lanl.\-gov/\-abs/\-cond-mat/\-9902318.

\noindent
Thirumalai,~D., and Kirkpatrick,~T.R., 1998, {\it Phys. Rev. B}, {\bf 38}, 
4881.

\noindent
Wu,~F., 1982, {\it Rev. Mod. Phys.}, {\bf 54}, 235.

\begin{figure}
\caption{
Qualitatively phase diagram for the PSG model in $3D$
as function of $s$. Solid lines mark real thermodynamic phase
transitions, while dotted 
line marks the vanishing Griffiths transition. Cross marks Potts
vanishing transition 
in the $s=1$ ($\pm J$ Ising SG) case. 
An analogous phase diagram holds for the PFF model
with a FF Ising phase at the place of the SG phase and without the G phase.
}
\label{fig1}
\end{figure}

\begin{figure}
\caption{
PSG for $s=2$: $\beta$ and $\tau$ parameters used to fit the nonlinear
susceptibility 
correlation function for $L\rightarrow \infty$ (from Ref.[breve]).
Circles are the
parameters for the form $iii)$, triangles for the form
$ii)$; squares are the integral correlation times. 
Where not shown the errors are smaller then symbols size.
Arrows show $k_BT_c/J=3.641$  and $k_BT_p/J=2.925\pm 0.075$.
}
\label{parpsg}
\end{figure}

\begin{figure}
\caption{
PFF with $s=2$: As in the previous figure.
Arrow shows $k_BT_p/J=2.76\pm0.05$.
}
\label{parpff}
\end{figure}

\begin{figure}
\caption{ 
PFF with $s=2$: as in the previous figure but for the Potts order
parameter's correlation function $f_M$. 
}
\label{pff_M}
\end{figure}

\newpage 

\begin{center}
\mbox{\epsfig{file=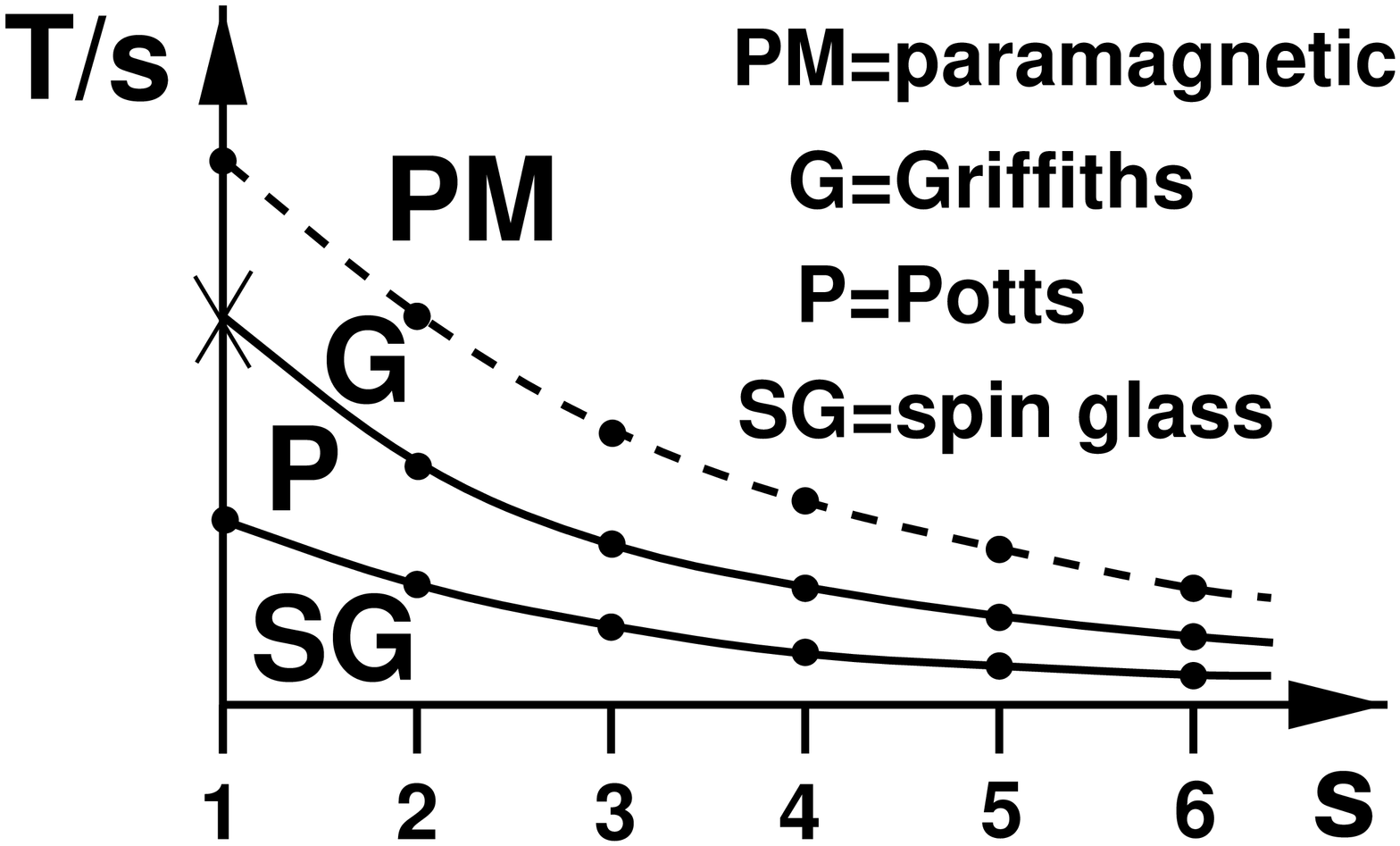,height=9cm,angle=270}}
\end{center}

\begin{center}
\mbox{\epsfig{file=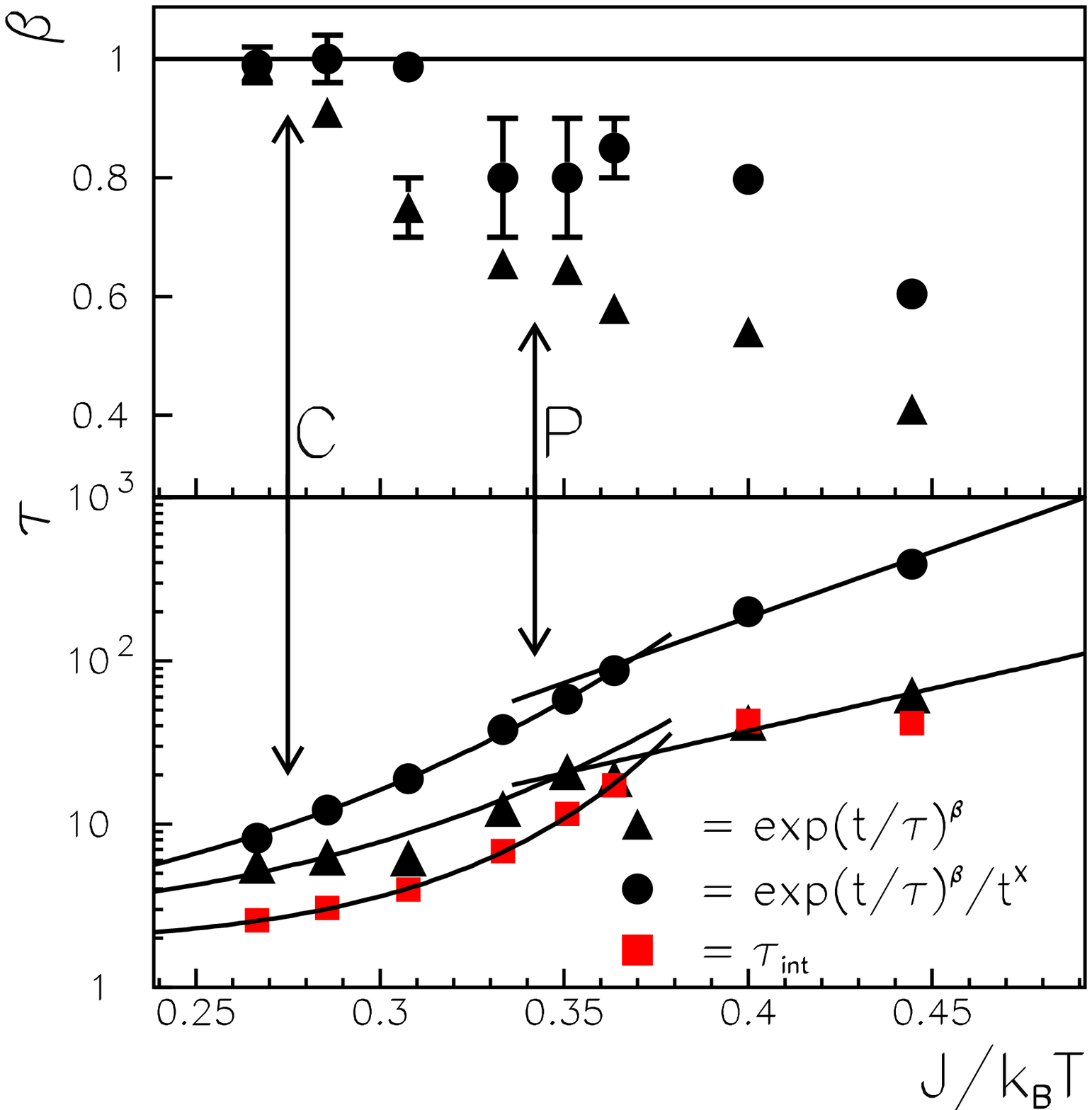,height=20cm}}
\end{center}


\begin{center}
\mbox{\epsfig{file=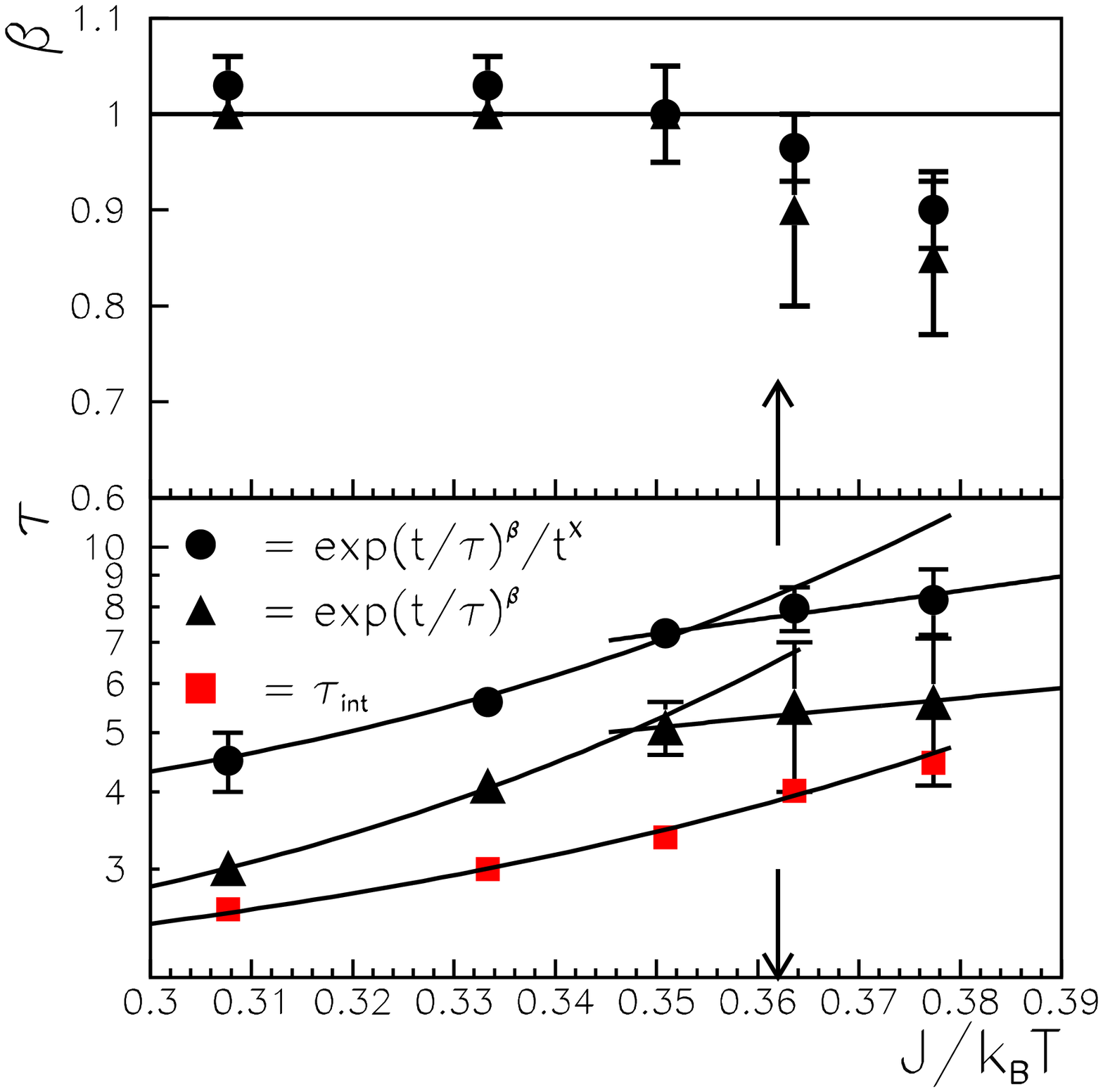,height=20cm}}
\end{center}


\begin{center}
\mbox{\epsfig{file=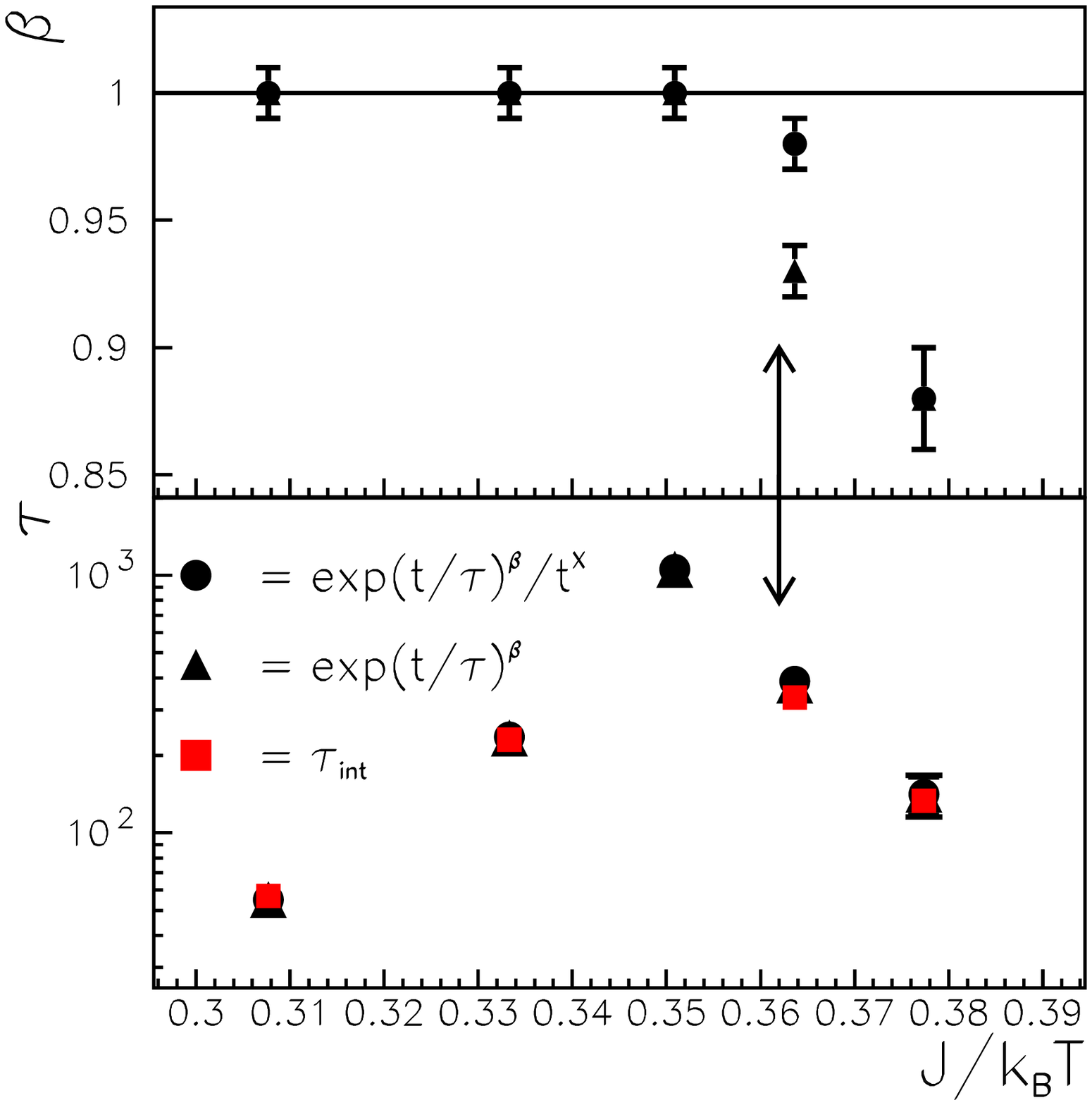,height=20cm}}
\end{center}


\end{document}